%% file: main.tex
\documentclass[apl,reprint]{revtex4-2}

\usepackage{xcolor}
\usepackage{siunitx}
\usepackage{graphicx}
\usepackage{amsmath}
\input{commands}

\begin{document}

\title{Microwave Phase Mapping and Angle-of-Arrival Detection Using Rydberg Atom-Based Electrometry} 

\author{Alexander Gill}
\author{Aaron Buikema}
\author{Stephen Sirisky}
\author{Hannah Clevenson}
\email[]{hclevenson@draper.com}
\affiliation{Charles Stark Draper Laboratory, Cambridge, MA 02139, USA}

\date{\today}

\begin{abstract}
  We present a method for simultaneously measuring the phase fronts of three or more RF fields using thermal Rydberg atoms.
  We demonstrate this method using an all-dielectric atomic electrometer acting in a heterodyne configuration to detect three single tone continuous-wave microwave signals propagating in free space.
  We show that this method can be used to measure the angle of arrival of each source.
  This method requires no calibration of the incident fields and makes no assumption about the form of the incident phase fronts.
  Because the sensor is minimally perturbing, it provides a direct probe of signal phase even in complicated environments, such as the near field of an antenna source or scattering surface.
  The measurements presented here are performed using a single scanned sensor, but this technique can trivially be extended to an array of sensors which would enable phase sensitive imaging and related applications.
\end{abstract}


\maketitle

\input{intro}
\input{method}
\input{experiment}
\input{discussion}
\input{conclusion}

\begin{acknowledgments}

The authors acknowledge the support of the Draper internal research and development program.

\end{acknowledgments}

\section*{Author Declarations}

\subsection*{Conflict of Interest}

The authors have no conflicts to disclose.

\subsection*{Author Contributions}

\textbf{Alexander Gill:} Conceptualization (equal); Data curation (equal); Formal analysis (lead); Methodology (equal); Administration (lead); Resources (equal); Software (equal); Supervision (lead); Validation (equal); Visualization (equal); Writing -- original draft (equal); Writing -- review \& editing (equal).
\textbf{Aaron Buikema:} Conceptualization (equal); Data curation (equal); Methodology (equal); Resources (equal); Software (equal); Validation (equal); Visualization (equal); Writing -- original draft (equal); Writing -- review \& editing (equal).
\textbf{Stephen Sirisky:} Conceptualization (equal); Data curation (equal); Methodology (equal); Resources (equal); Writing -- original draft (supporting); Writing -- review \& editing (equal).
\textbf{Hannah Clevenson:} Conceptualization (equal); Funding acquisition (lead); Methodology (equal); Resources (equal); Writing -- original draft (supporting); Writing -- review \& editing (equal).

\section*{Data Availability Statement}

The data that support the findings of this study are available from the corresponding author upon reasonable request.

\bibliography{rydberg}

\appendix
\input{appendix}

\end{document}

%% file: commands.tex
\newcommand{\spectroscopynotation}[3]{${#1}\,{#2}_{#3}$}

%% file: intro.tex
\section{Introduction}\label{sec:intro}

Rydberg atom-based radio frequency electric field detection is a rapidly developing technology area~\cite{Fancher2021,Meyer2020}.
There is growing interest in developing these devices for use as low-SWAP, high-sensitivity replacements for traditional antennas and front-end receivers~\cite{Meyer2021,Jing2020,Holloway2019}.
These sensors can be configured for rapid tuning over extremely wide frequency ranges (spanning MHz to THz)~\cite{Downes2020}.
Due to the use of optical detection methods, they can be built with non-conductive materials that negligibly perturb the field being measured, allowing for the development of non-interacting sensor arrays.
This measurement would utilize the electrically small, omnidirectional and non-interacting nature of the sensor as a way to perform a direct spatial measurement without requiring phase calibration.
Using atom-based sensors along with software defined radios (SDRs) for RF detection allows for an unprecedented departure from the inverse correlation between frequency and antenna size.

Angle of arrival (AoA) measurements can be used to geo-locate signals from a variety of sources including electromagnetics and acoustics.
From locating pirate radio stations to calls to emergency services, this technique is used broadly across many fields. 
Typically, these measurements require rotating antennas or moving a single antenna, or antenna arrays.
It has been shown that AoA determination from sub-wavelength phase difference measurements is feasible using Rydberg atom-based sensors~\cite{Robinson2021}. That demonstration, however, was contingent on the assumption that the incoming source and LO signals were pure plane waves.
Furthermore, the utilization of only a single pair of RF signals is only sufficient to determine the relative angle of those two propagating waves, leaving the azimuthal coordinate of the AoA undetermined.
The method presented here is fully general, and with a large enough sampling region can fully disambiguate incident angles in three dimensions.

We simulate the behavior of an array of these sensors and demonstrate a) a table-top measurement of the non-uniform phase front of an incident RF field, and b) an estimate of the angle of arrival of each incoming field.
No calibration of any given field beforehand is required; the individual phase of each source can be extracted independently.
Further, we show that it is robust across a range of incident frequencies.

%% file: method.tex
\section{Method}\label{sec:method}

\begin{figure}
    \centering
    \includegraphics[width=0.6\columnwidth]{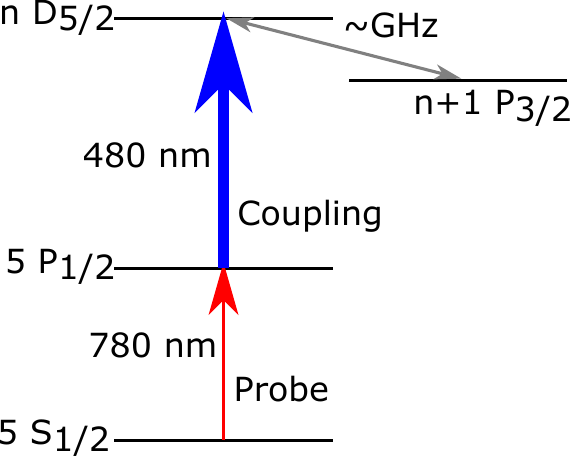}
    \caption{Rb energy level structure. A probe laser couples the first pair of states, while a coupling laser couples to a Rydberg state. An RF field couples together two Rydberg states.}
    \label{fig:Rb_levels}
\end{figure}

Rydberg atom-based sensors utilize the large polarizability of highly-excited atoms to measure incident electric fields.
Electromagnetically-induced transparency provides an optical readout method for measuring the change in energy of these Rydberg states (via the Autler-Townes splitting or AC Stark effect) due to an incident electric field~\cite{Holloway2014}.
Figure~\ref{fig:Rb_levels} shows a typical set of energy levels in rubidium used for this type of measurement.
With appropriate probe and coupling lasers present, an incident electric field will change the susceptibility of the vapor as seen by the probe laser.

Unlike traditional sensors used in antenna characterization (such as e.g. an open-ended waveguide probe), atomic E-field sensors of standard construction~\cite{Sedlacek2012,Holloway2014} are sensitive only to the E-field envelope and have no inherent phase sensitivity.
Sensitivity to relative phase between multiple coherent electric fields of similar frequency may be achieved by beating these fields together at the sensor in a heterdyne configuration~\cite{Simons2019,Jing2020}.
Using a properly arranged experimental setup, we exploit this effect to map out the phase fronts of the incident fields.
A full derivation of this technique is provided in Appendix~\ref{sec:full_derivation}; we review the salient details here.

We consider the case of three incident free-space oscillating electric fields at angular frequencies $\omega_i$ near the relevant atomic resonance and (position-dependent) phase $\phi_i$ for $i=0,1,2$:
\begin{equation}
    \mathcal{E}_i(\mathbf{x},t) = E_{i}(\mathbf{x}) \cos \left(\omega_i t + \phi_i(\mathbf{x})\right)
\end{equation}
where $E_{i}(\mathbf{x})$ is the amplitude of the incident beam.
Note that this term can also be position dependent.
For notational simplicity, we assume all fields have the same polarization, and we will omit the explicit time and position dependence.

A given atomic sensor head is located at a point in space within the radiation pattern of all three fields.
The laser sources and optical detector are located remotely from the test area to avoid perturbing the fields.
Due to the nonlinear effect of the atoms, this will perform a heterodyne measurement.
The resulting measured beat note will have frequency $\omega_{ij}=\omega_i-\omega_j$ and phase $\phi_{ij}=\phi_i-\phi_j$; this assumes $\omega_{ij}$ is within the linewidth of this transition (generally $\lesssim \SI{1}{MHz}$).
This beat note phase can then be measured at different points in space, either simultaneously using an array of sensors, or by moving a single sensor through space.

If the spatial phase dependence of one of the incident fields ($\phi_{i}$) is known precisely from prior calibration, then determining the wavefront map of the other incident fields is trivial: we can easily determine $\phi_j$ by measuring $\phi_{i,j}$ at a number of points.
This calibration may be difficult or impractical to achieve in many contexts, for example, in the near-field regime of any source antenna.
For that reason, it is desirable to be able to measure the phase fronts with arbitrary incident fields.

With three incident fields, one can disambiguate the relative RF phase of each incident field, which allows mapping out the phase fronts \emph{even when no incident field has been calibrated}.
Note that the fields must be non-degenerate, i.e., the phase difference between fields must have some spatial dependence.

Having determined the phase of the signal from each RF source over an array of spatial positions in the measurement region, we may numerically calculate the wave vector at each position using
\begin{equation}
  \label{eq:3}
  \mathbf{k}(\mathbf{x}) = \nabla \phi(\mathbf{x})\,.
\end{equation}
We may then calculate $\vartheta$ and $\varphi$, the local polar and azimuthal incidence angles of the field at each point, which we define to indicate the direction counter to $\mathbf{k}(\mathbf{x})$ (i.e. pointing back towards the source).

%% file: experiment.tex
\section{Experimental Demonstration}\label{sec:experiment}

\begin{figure}
    \centering
    \includegraphics[width=0.9\columnwidth]{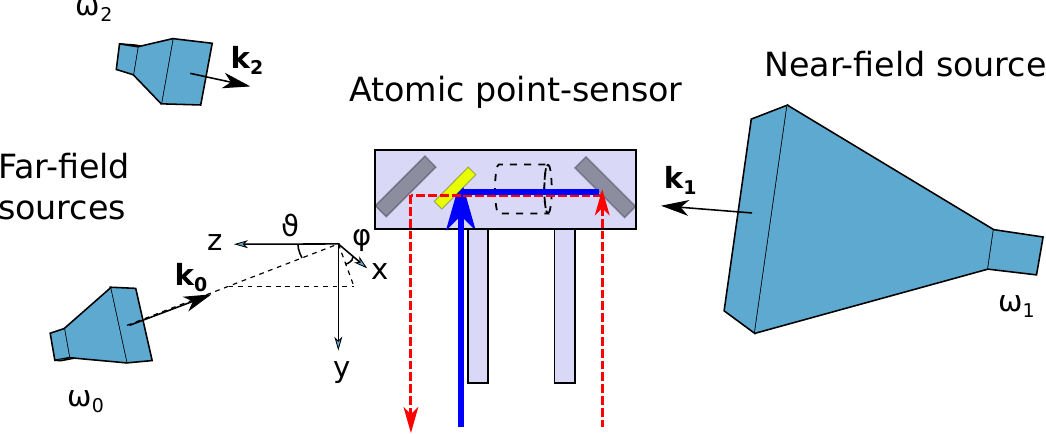}
    \caption{Schematic of experimental setup (not to scale, antenna positions notional).
    The Rb vapor cell sits inside a 3D-printed polycarbonate enclosure above the optical table.
    Lasers enter from below via optical fibers and are overlapped with a dichroic mirror.
    The entire setup sits on an X-Y motion stage.
    The transmission of the \SI{780}{nm} probe is measured as the sensor is moved.}
    \label{fig:setup}
\end{figure}

Figure~\ref{fig:setup} shows the tabletop free-space setup used in this demonstration.
Three fixed antennas are directed toward a Rb vapor cell.
The antennas are variously positioned at distances from the cell ranging approximately \SI{0.4}{m} to \SI{2}{m}.
The vapor cell and basic steering mirrors sit inside a 3D-printed polycarbonate enclosure, which is supported above the table and is approximately \SI{25}{mm} long.
The sensor enclosure is mounted above an X-Y motion stage to spatially sample the incident RF fields.
The entire sensor structure contains no conductors and thus minimally perturbs the incident RF fields.
Frequency-stabilized probe (\SI{780}{nm}) and coupling (\SI{480}{nm}) lasers enter from below via optical fibers and counter-propagate through the cell, and the transmitted probe beam power is measured vs sensor position.
The powers of the probe and coupling beams were measured to be \SI{30}{\micro\watt} and \SI{62}{\milli\watt}, while the beam diameters are estimated to be \SI{1.2}{\milli\meter} and  \SI{1.0}{\milli\meter} at the sensor, respectively.

Figure~\ref{fig:Rb_levels} shows the Rb energy levels for this system.
We use the \spectroscopynotation{5}{S}{1/2}$\rightarrow$\spectroscopynotation{5}{P}{3/2} transition of ${}^{85}$Rb at \SI{780}{nm}, while a tunable \SI{480}{nm} laser is used to couple to Rydberg states.
We perform these measurements at two RF frequencies: $\omega_{\text{base}}=2\pi\times$\SI{12.0073}{GHz}, which is resonant with the \spectroscopynotation{56}{D}{5/2}$\rightarrow$\spectroscopynotation{57}{P}{3/2} transition, and $\omega_{\text{base}}=2\pi\times$\SI{7.262}{GHz}, which is resonant with the \spectroscopynotation{66}{D}{5/2}$\rightarrow$\spectroscopynotation{67}{P}{3/2} transition.

A polarization spectroscopy lock~\cite{Harris2006,Carr2012} to a separate vapor cell is used to stabilize both lasers to the relevant EIT transitions.
This scheme allows for a wide capture range, simplified electronics, and does not require a separate modulation source.

The antennas are connected to individual reference-locked signal generators operating at angular frequencies $\omega_{i} = \omega_{\text{base}} + \omega_{\text{offset}}$, where $\omega_{\text{offset}} = 2\pi\times \{38, 26, 0\}~\si{kHz}$ for $i=0,1,2$.
These offset frequencies were chosen to avoid beating with existing spectral lines, including harmonics of the desired beat note frequencies.
This results in the appearance of beat signals in the probe transmission at frequencies $\omega_{10} = 2\pi\times\SI{-12}{kHz}$, $\omega_{21} = 2\pi\times\SI{-26}{kHz}$, and $\omega_{02} = 2\pi\times\SI{38}{kHz}$.
As in a traditional heterodyne measurement, we measure the phases of each of these beat signals using lock-in detection.
The sensor position is scanned over a two dimensional grid spanning \SI{36}{cm} by \SI{36}{cm}, and the phases are recorded as a function of grid position.
The electronics for the beat note measurement are also phase-locked to the same reference as the signal generators, ensuring phase stability for the duration of the experiment.

The beat signal phase data is processed as described in Appendix~\ref{sec:full_derivation} to extract the phase of the signal from each source antenna at each point in the grid.
Since our phase data is only determined in a single measurement plane, the gradient of $\phi$ only directly gives us $\mathbf{k}_\parallel = (k_{x}, k_{y})$, the component of $\mathbf{k}$ in the $XY$-plane.
From this we can determine the azimuthal incidence angle
\begin{equation}
  \label{eq:1}
  \varphi = \mathrm{arctan2}(-k_{y}, -k_{x})
\end{equation}
directly using the two-argument arc-tangent function.
Without additional phase gradient data out of the plane, the polar incidence angle $\vartheta$ cannot be determined unambiguously, but it can be determined up to a reflection across the $XY$-plane.
To determine the polar angle we use the fact that the total wave vector magnitude $k=\omega/c$ is known.
The polar angle is then calculated as
\begin{equation}
  \label{eq:2}
  \vartheta = \arcsin(|k_{\parallel}|/k)\,.
\end{equation}
Finally, for any source known to be  positioned below the measurement plane, we manually correct the ambiguity in the polar angle by reflecting it over this plane.

%% file: discussion.tex
\section{Results and Discussion}\label{sec:discussion}

\begin{figure*}
    \centering
    \includegraphics[width=0.8\textwidth]{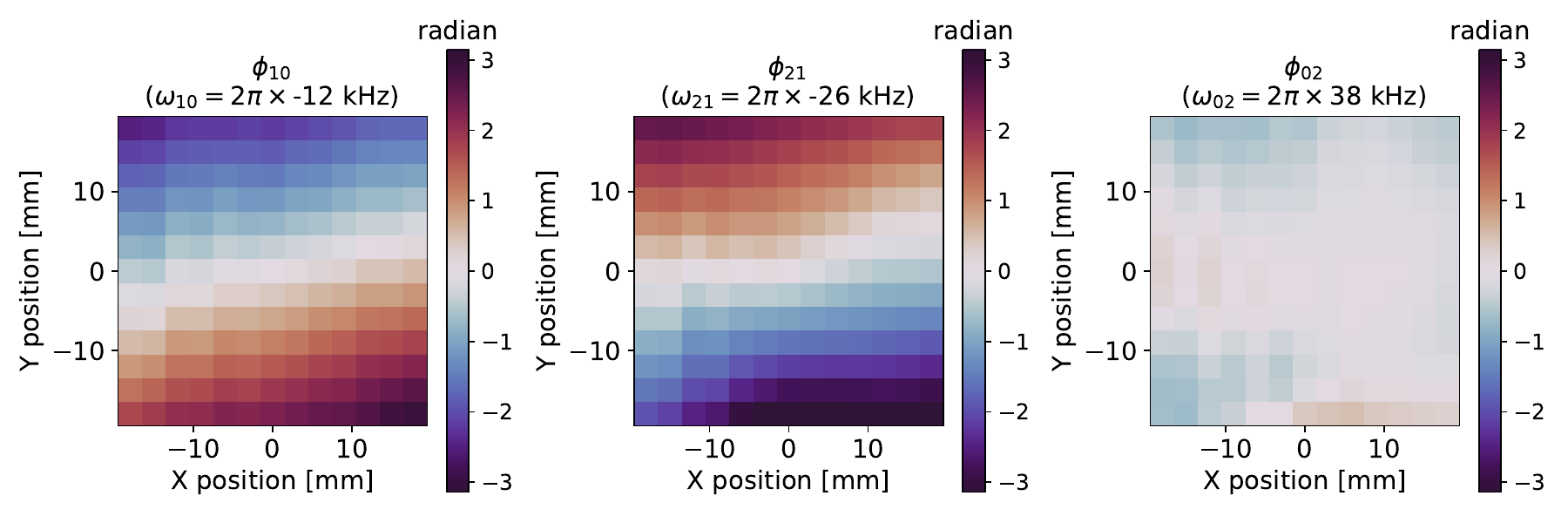}
    \caption{Measured phase of beat signals vs. position at \SI{7}{GHz} carrier frequency.}
    \label{fig:beat_note_phase}
\end{figure*}
The demodulated beat signal phases are shown in Figure~\ref{fig:beat_note_phase} as a function of position within the measurement grid.
Since it is only the relative phase between sensor positions that is physically relevant, we have arbitrarily chosen to zero the phases at the origin, which is the center position in the grid.
Each source frequency $\omega_{i}$ is uniquely associated with a single antenna, but because each beat signal derives from a pair of sources, the spatially varying phase of each antenna signal cannot be observed directly from the data shown in Figure~\ref{fig:beat_note_phase}.
We note that the phase progression of each beat signal across the measurement region appears relatively uniform, with apparent absence of large scale structure.
This can be interpreted as an indication that the wavefronts of all the observed antenna signals are nearly planar, which is an expected result given that measurement area is relatively small, and none of the antennas are in the extreme near field.

\begin{figure*}
    \centering
    \includegraphics[width=0.8\textwidth]{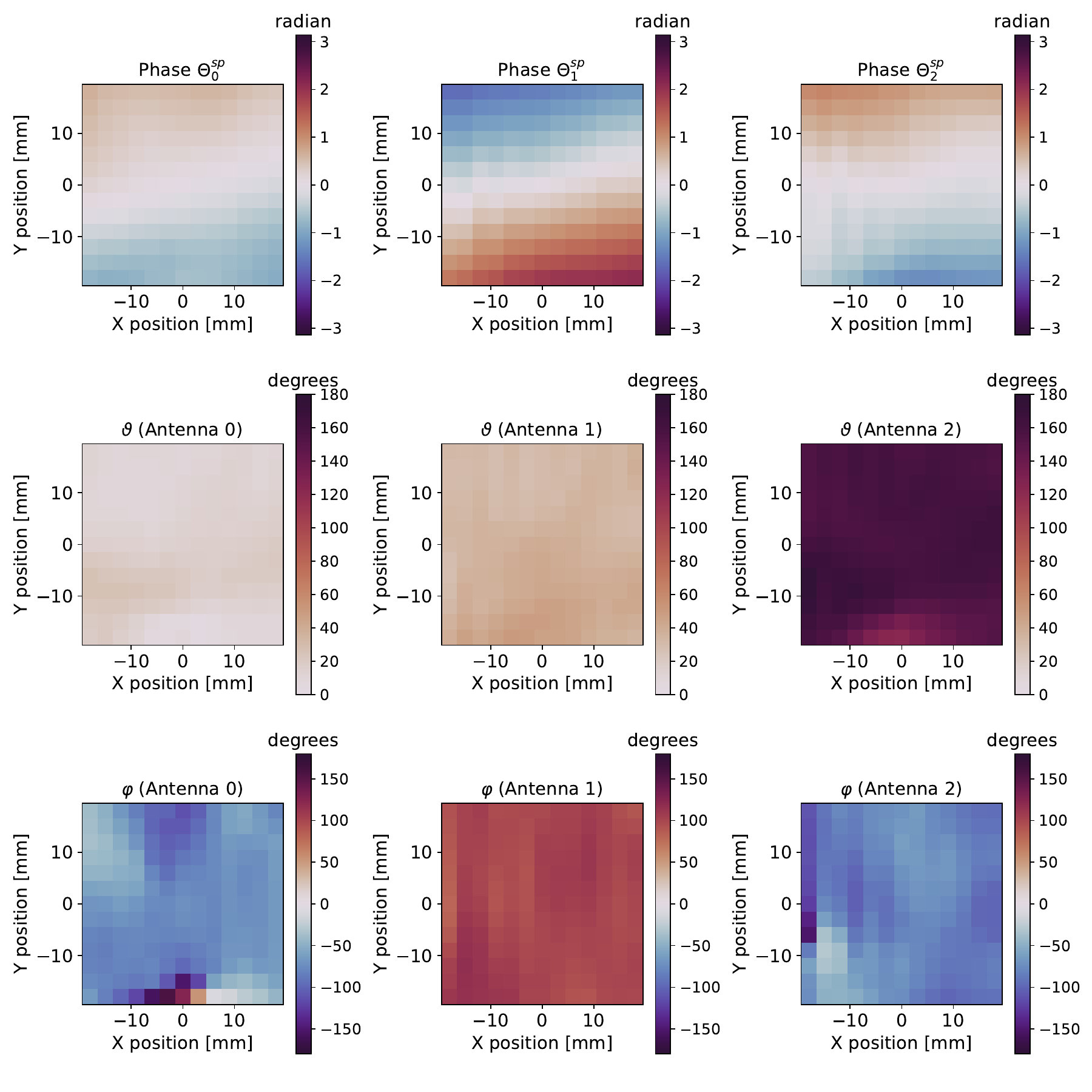}
    \caption{(Top row) Measured source phase ($\theta^{\textrm{sp}}$) using the $\omega_{\text{base}} = 2\pi\times\SI{7}{GHz}$ data set at each grid position.
      The second and third rows give the estimated polar and azimuthal angles to the source (AoA) at each point based on the calculated phase gradient.
      The columns are grouped according to source offset frequency, with each frequency emanating from a unique antenna.}
    \label{fig:source_gradients}
\end{figure*}
To separate the phase information attributable to each antenna, we apply the transformation of eq.~\eqref{eq:spatial_phase_matrix}.
This yields the three spatially dependent phases $\theta^{\textrm{sp}}_{i}$ shown in the top row of Figure~\ref{fig:source_gradients}.
Computing the gradient, eq.~(\ref{eq:3}), on a point-wise basis yields the local wave vector $\mathbf{k(\mathbf{x})}$, from which we calculate the local incidence angles $\vartheta$ and $\varphi$ via eqs.~(\ref{eq:1}) and (\ref{eq:2}).
These angles are shown in the second and third rows of Figure~\ref{fig:source_gradients}.
The uniformity of the observed incidence angles over the measurement region confirms that the phase fronts there are approximately planar.
We note that we make no assumption of planar wavefronts in this analysis, and larger phase anomalies would be expected in the presence of greater distortion.
For this experiment we intentionally took care to avoid the use of highly absorbing or reflecting materials near the vapor cell in order to limit distortion.

\begin{figure*}
    \centering
    \includegraphics[width=0.8\textwidth]{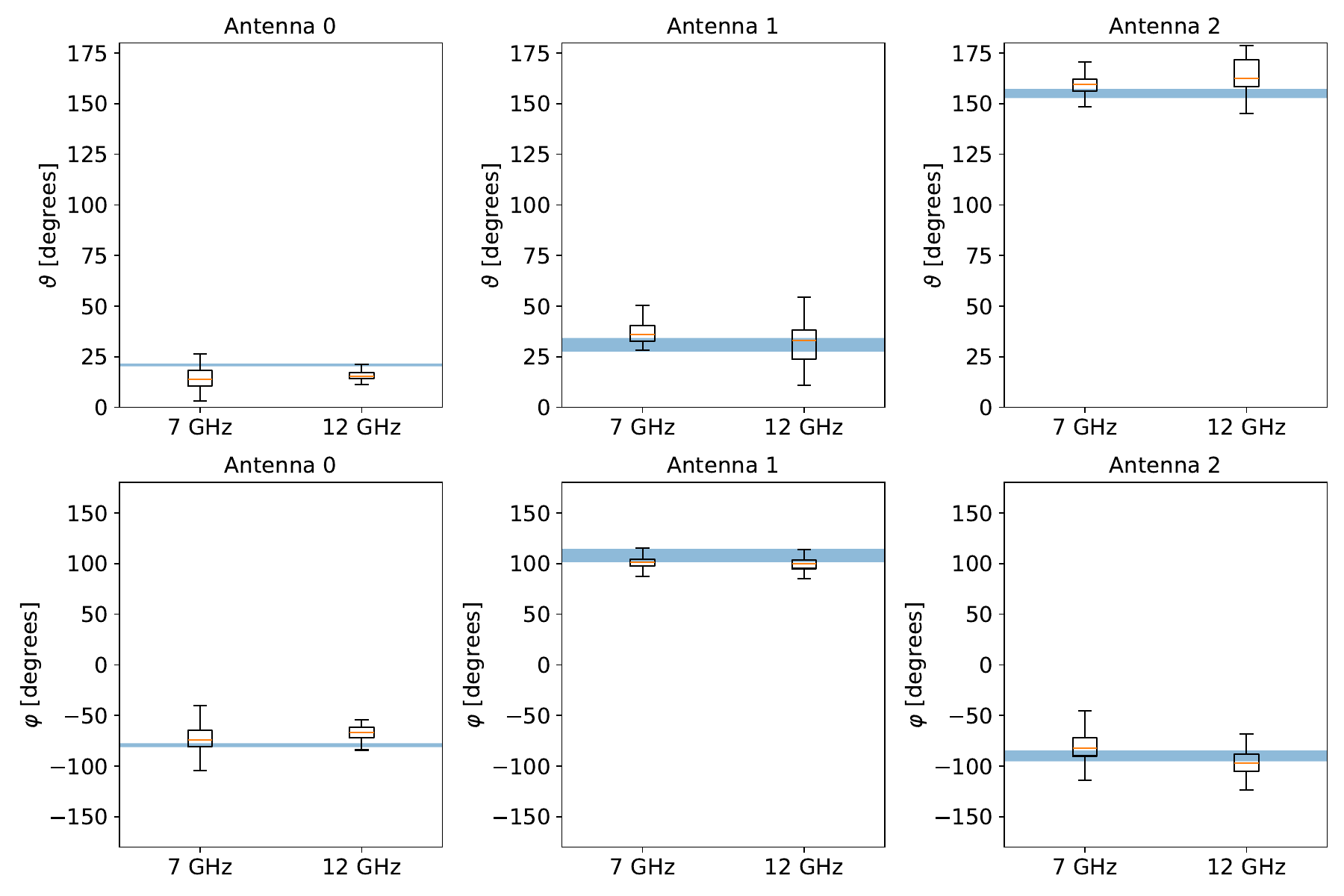}
    \caption{Comparison of estimated polar and azimuthal angles ($\vartheta$ and $\varphi$) of the signals from each source at the two measurement frequencies, $\omega_{\text{base}}/2\pi$ = \SI{7}{GHz} and \SI{12}{GHz}.
      The columns are grouped according to source offset frequency, with each frequency emanating from a unique antenna.
      The antenna positions are identical for both the \SI{7}{GHz} and \SI{12}{GHz} data sets.
      Each grid position is treated as an independent measurement with box plots showing the distribution of angles over all measurement positions with whisker length 1.5 times the inter-quartile range.
      The shaded region indicates the estimated direction from the center of the grid to the center of each antenna (1-$\sigma$ confidence band) based on position measurements of each.}
    \label{fig:source_AoA}
\end{figure*}

Finally in Figure~\ref{fig:source_AoA} we plot the statistical distributions of $\vartheta$ and $\varphi$ for over all points in the measurement region.
An identical analysis was performed for both data sets taken using the \SI{7}{GHz} and the \SI{12}{GHz} carrier frequencies, and the results are compared in the figure.
These distributions show excellent agreement in the predicted incidence angles of each antenna signal.
Recall that the measurements taken at the two carrier frequencies are fundamentally independent experiments, utilizing different Rydberg levels.
The agreement of the predicted source angles between the two experiments is evidence of the basic soundness of the directional sensing technique.

If the phase fronts were perfectly planar across the measurement region, we would expect the measured incidence angles to match those of a unit vector pointing from the center of the measurement region toward the source.
We approximately determined these pointing angles in the lab setup based on measurements of the Cartesian positions of the center of each antenna aperture relative to the center of the measurement region.
These independently measured angles are represented in Figure~\ref{fig:source_AoA} by the blue shaded regions.
The width of each band is twice the 1-$\sigma$ uncertainty in each angle deriving from the position measurement uncertainty.
The excellent match between the atomically measured incidence angles and the expected angles is further evidence of the soundness of the technique.

%% file: conclusion.tex
\section{Conclusion}\label{sec:conclusion}

We present a scheme for direct measurement of the phase fronts of RF electric fields using an atomic sensor as the sensing element and experimentally demonstrate the approach using a single scanned sensor in a laboratory environment.
Angle-of-arrival measurements based on this scheme are in excellent agreement with expected results.
All-dielectric Rydberg atom-based electrometers offer an ideal sensor platform for realizing this scheme, because they minimally perturb the incident fields and receive omnidirectionally.
These sensors have the added benefit of sensitivity across a wide spectrum.
This method can be extended to an entire array of atomic sensors, allowing for instantaneous readout and eliminating the need to scan sensor positions.
We expect that this method has applicability to any field sensor that operates on the principle of interference between propagating waves for relative phase determination and can be applied in other regions of the spectrum accessible to atomic electrometers and magnetometers.
In future work we intend to explore the limits of precision and accuracy of this method in a more controlled test environment.
Additionally, by using a sensor system with a well calibrated amplitude response, it should be possible to determine signal amplitude in addition to signal phase, providing a complete and instantaneous picture of the state of an RF field.
Such a total field sensor would find use in numerous applications and techniques such as phase sensitive imaging, synthetic aperture radar, and near-field antenna characterization.

%% file: appendix.tex
\section{Derivation of method}\label{sec:full_derivation}

We provide a derivation of the extraction of individual field spatial phase delays.

\subsection{Beat note detection}
\label{sec:beat_note_detection}
When irradiated by a pair of incident RF fields near a resonance
\begin{eqnarray}
  \label{eq:30}
  \mathcal{E}_1(\mathbf{x},t) & = & E_{1}(\mathbf{x}) \cos\left(\omega_1 t + \phi_1(\mathbf{x})\right) \\
  \label{eq:31}
  \mathcal{E}_2(\mathbf{x},t) & = & E_{2}(\mathbf{x}) \cos\left(\omega_2 t + \phi_2(\mathbf{x})\right)\,,
\end{eqnarray}
the atomic sensor measures the (rectified) beat note amplitude
\begin{equation}
  A = \sqrt{E_1^2 + E_2^2 + 2 E_1 E_2 \cos\left(\omega_{21} t+\phi_{21}\right)},
\end{equation}
where $\phi_{21} \equiv \phi_{2} - \phi_{1}$ and $\omega_{21} \equiv \omega_{2} - \omega_{1}$.
The (temporal) Fourier component at $\omega=\omega_{21}$, \emph{i.e.}, what is measured in a lock-in detection method, will have an amplitude that is a complicated function of the incident field amplitude.
The phase of this component, however, is always $\phi_{21}$.
For this reason, the phase difference between incidence electric fields can always be correctly extracted even when not in the traditional LO/small-signal regime.

\subsection{Three incident fields}
\label{sec:interf-field-meas}
Consider the case of three phase-coherent fields each with slightly different frequencies all incident on a single atomic sensor.
No assumptions are made about the spatial phase pattern of each field.
The output of the sensor can be read out at each unique beat frequency $\omega_{ij}$ with an associated phase $\phi_{ij}$.

The three demodulated phase readings $\phi_{ij}$ produce the following system of equations:
\begin{eqnarray}
  \label{eq:34}
  \phi_{10} & = & \phi_{1} - \phi_{0} + \Theta^{\textrm{post}}_{10} + \psi_{10}\\
  \label{eq:35}
  \phi_{21} & = & \phi_{2} - \phi_{1} + \Theta^{\textrm{post}}_{21} + \psi_{21} \\
  \label{eq:36}
  \phi_{02} & = & \phi_{0} - \phi_{2} + \Theta^{\textrm{post}}_{02} + \psi_{02}\,.
\end{eqnarray}
$\phi_{i}$ with $i \in \{0,1,2\}$ is the phase associated with each individual field, where the first difference $\phi_i-\phi_j$ is the intrinsic beat note phase.
$\Theta^{\textrm{post}}_{ij}$ is shorthand for $\Theta^{\textrm{post}}(\omega_{ij})$, the phase response of the detection path (e.g. photodetector and other electronic delays).
Generally $\Theta^{\textrm{post}}(\omega)$ will be constant across beat frequencies of interest, or it can be measured independently and removed.
$\psi_{ij}$ is an arbitrary overall correction that allows us to define $\phi_{ij}=0$ at some $t=0, \mathbf{x}=0$.

The individual field phases can be decomposed as
\begin{equation}\label{eq:field_phase}
  \phi_{i} \equiv \phi(\mathbf{x}, \omega_{i}) = \Theta^{\textrm{pre}}_{i}(\omega_{i}) + \theta^{\textrm{sp}}_{i}(\mathbf{x}, \omega_{i}),
\end{equation}
where $\Theta^{\textrm{pre}}_{i}(\omega_{i})$ encompasses the phase offset of source $i$ and phase delays imparted by the signal path up to the feed antenna.
Finally, $\theta^{\textrm{sp}}_{i}(\mathbf{x}, \omega_{i})$ is the spatial phase delay of field $i$ at frequency $\omega_{i}$, or the phase of the antenna pattern.

We note that $\theta^{\textrm{sp}}_{i}(\mathbf{x}, \omega_{i})$ is an unknown function of space and signal frequency, and we make no assumptions about its form.
For traveling waves the dependence on $\omega$ comes in through both the wavevector $\mathbf{k}$ with magnitude $k=\omega/c$ and any frequency dependence of the antenna pattern.
For brevity, we will drop the explicit dependence on $\omega$ going forward.
For the simplest example of a plane wave, it will take the form
\begin{equation}
  \label{eq:plane_wave_phase}
  \theta^{\textrm{sp}}(\mathbf{x}) = \mathbf{k} \cdot \mathbf{x} + \theta^{\textrm{sp}}(0)\,.
\end{equation}

We choose a spatial origin point to serve as a global phase zero.
We will set each demodulated phase to zero at the origin point by ad hoc adjustment of the reference phases $\psi_{ij}$ with the sensor positioned at the origin.
We note that no prior calibration of $\Theta^{\textrm{pre}}_{i}$ and $\Theta^{\textrm{post}}$ is necessary as these have no spatial dependence, so their contributions to phase can be canceled at the origin by the tuning of $\psi_{ij}$.

We now have reduced our system of equations to only the spatially dependent terms
\begin{eqnarray}
  \label{eq:40}
  \phi_{10}(\mathbf{x}) & = & \theta^{\textrm{sp}}_{1}(\mathbf{x}) - \theta^{\textrm{sp}}_{0}(\mathbf{x}) \\
  \label{eq:41}
  \phi_{21}(\mathbf{x}) & = & \theta^{\textrm{sp}}_{2}(\mathbf{x}) - \theta^{\textrm{sp}}_{1}(\mathbf{x}) \\
  \label{eq:42}
  \phi_{02}(\mathbf{x}) & = & \theta^{\textrm{sp}}_{0}(\mathbf{x}) - \theta^{\textrm{sp}}_{2}(\mathbf{x})\,,
\end{eqnarray}
or in matrix form
\begin{equation}
  \label{eq:43}
  \begin{pmatrix}
    \phi_{10}\\\phi_{21}\\\phi_{02}
  \end{pmatrix}
  =
  \begin{pmatrix}
    -1 & 1 & 0 \\
    0 & -1 & 1 \\
    1 & 0 & -1
  \end{pmatrix}
  \begin{pmatrix}
    \theta^{\textrm{sp}}_{0}\\\theta^{\textrm{sp}}_{1}\\\theta^{\textrm{sp}}_{2}
  \end{pmatrix}\,.
\end{equation}
The matrix is singular with rank 2, therefore an infinite number of solutions exist.
We can obtain the family of solutions by taking the Moore-Penrose pseudoinverse, yielding
\begin{equation}
  \label{eq:spatial_phase_matrix}
  \begin{pmatrix}
    \theta^{\textrm{sp}}_{0}\\\theta^{\textrm{sp}}_{1}\\\theta^{\textrm{sp}}_{2}
  \end{pmatrix}
  = \frac{1}{3}
  \begin{pmatrix}
    -1 & 0 & 1 \\
    1 & -1 & 0 \\
    0 & 1 & -1
  \end{pmatrix}
  \begin{pmatrix}
    \phi_{10}\\\phi_{21}\\\phi_{02}
  \end{pmatrix}
  +
  \begin{pmatrix}
    1 \\ 1 \\ 1
  \end{pmatrix}
  \psi_{\textrm{arb}}\,.
\end{equation}
The global phase $\psi_{arb}$ is a constant that can take any value and may be set arbitrarily to zero.
This has the effect of setting the spatial phases of all three fields to zero at the origin.
We now have a method of measuring the spatial phases of all three fields at any point in space by measuring the phases of the beat notes at any point and performing the linear combination of eq.~\eqref{eq:spatial_phase_matrix}.

\subsection{Significance of using three fields}
\label{sec:sign-using-three}
Assume we use only two incident fields, with frequencies $\omega_{0}$ and $\omega_{1}$ launched into free-space using separate antennas.
We may detect a demodulated beat signal at only one frequency $\omega_{10}$.
Again, $\psi_{10}$ is the arbitrary reference phase introduced at the demodulation step. Our demodulated phase is then given by the single equation
\begin{equation}
  \label{eq:45}
  \begin{split}
    \phi_{10} & = \theta^{\textrm{sp}}_{1}(\mathbf{x}) - \theta^{\textrm{sp}}_{0}(\mathbf{x}) \\
    & + \Theta^{\textrm{pre}}_{1}(\omega_{1}) - \Theta^{\textrm{pre}}_{0}(\omega_{0}) + \Theta^{\textrm{post}}(\omega_{10}) + \psi_{10}\,.
  \end{split}
\end{equation}
Again, we may choose a spatial origin point and set the demodulation phase such that the phase of the beat signal is zero at the origin.
We are left with a measurement of the spatially-dependent difference phase between the two fields $\phi_{10} = \theta^{\textrm{sp}}_{1} - \theta^{\textrm{sp}}_{0}$.
This is a useful measurement only if one of the two fields has a known spatial phase pattern.
This is not likely practical in an application such as antenna characterization, particularly when the antenna sources may be interacting in the near field, rendering prior calibration unreliable.

Now consider the case where we have four or more fields.
Assume we have $N$ fields, launched into free-space using separate antennas, and again we choose the frequencies $\omega_{i}$ such that all the beat frequencies $\omega_{ij}$ are unique (and higher harmonics do not overlap), and we zero each demodulation phase at the origin.
We now have up to $\binom{N}{2}$ unique beat signals at our disposal.
There are a number of ways to construct a system of equations of rank $N-1$ from these demodulation signals.
Consider the case of $N=4$ fields (with up to six beat signals).
Generalizing from eq.~\eqref{eq:43}, we may choose the rank 3 system
\begin{equation}
  \label{eq:46}
  \begin{pmatrix}
    \phi_{10}\\\phi_{21}\\\phi_{32}\\\phi_{03}
  \end{pmatrix}
  =
  \begin{pmatrix}
    -1 & 1 & 0 & 0 \\
    0 & -1 & 1 & 0 \\
    0 & 0  & -1 & 1 \\
    1 & 0 & 0 & -1
  \end{pmatrix}
  \begin{pmatrix}
    \theta^{\textrm{sp}}_{0}\\\theta^{\textrm{sp}}_{1}\\\theta^{\textrm{sp}}_{2}\\\theta^{\textrm{sp}}_{3}
  \end{pmatrix}
\end{equation}
with solution
\begin{equation}
  \label{eq:47}
  \begin{pmatrix}
    \theta^{\textrm{sp}}_{0}\\\theta^{\textrm{sp}}_{1}\\\theta^{\textrm{sp}}_{2}\\\theta^{\textrm{sp}}_{3}
  \end{pmatrix}
  = \frac{1}{8}
  \begin{pmatrix}
    -3 & -1 & 1 & 3 \\
    3 & -3 & -1 & 1 \\
    1 & 3 & -3 & -1 \\
    -1 & 1 & 3 & -3
  \end{pmatrix}
  \begin{pmatrix}
    \phi_{10}\\\phi_{21}\\\phi_{32}\\\phi_{03}
  \end{pmatrix}
  +
  \begin{pmatrix}
    1 \\ 1 \\ 1 \\ 1
  \end{pmatrix}
  \psi_{\textrm{arb}}\,,
\end{equation}
where $\psi_{\textrm{arb}}$ is again an arbitrary global phase that we will set to zero.
Alternatively, we may utilize all six demodulation signals to obtain the rank 3 system
\begin{equation}
  \label{eq:48}
  \begin{pmatrix}
    \phi_{10}\\\phi_{20}\\\phi_{30}\\\phi_{21}\\\phi_{31}\\\phi_{32}
  \end{pmatrix}
  =
  \begin{pmatrix}
    -1 & 1 & 0 & 0 \\
    -1 & 0 & 1 & 0 \\
    -1 & 0 & 0 & 1 \\
    0 & -1 & 1 & 0 \\
    0 & -1 & 0 & 1 \\
    0 & 0 & -1 & 1
  \end{pmatrix}
  \begin{pmatrix}
    \theta^{\textrm{sp}}_{0}\\\theta^{\textrm{sp}}_{1}\\\theta^{\textrm{sp}}_{2}\\\theta^{\textrm{sp}}_{3}
  \end{pmatrix}
\end{equation}
with solution
\begin{equation}
  \label{eq:49}
  \begin{split}
  \begin{pmatrix}
    \theta^{\textrm{sp}}_{0}\\\theta^{\textrm{sp}}_{1}\\\theta^{\textrm{sp}}_{2}\\\theta^{\textrm{sp}}_{3}
  \end{pmatrix}
  & = \frac{1}{4}
  \begin{pmatrix}
    -1 & -1 & -1 & 0 & 0 & 0 \\
    1 & 0 & 0 & -1 & -1 & 0 \\
    0 & 1 & 0 & 1 & 0 & -1 \\
    0 & 0 & 1 & 0 & 1 & 1
  \end{pmatrix}
  \begin{pmatrix}
    \phi_{10}\\\phi_{20}\\\phi_{30}\\\phi_{21}\\\phi_{31}\\\phi_{32}
  \end{pmatrix} \\
  & +
  \begin{pmatrix}
    1 \\ 1 \\ 1 \\ 1
  \end{pmatrix}
  \psi_{\textrm{arb}}\,,
\end{split}
\end{equation}
where we arbitrarily set $\psi_{\textrm{arb}}$ to zero.
Both of these approaches should yield identical results, though the latter approach may be less sensitive to noise, since it takes into account more experimental data.

In conclusion, $N=3$ fields is special in that it is the fewest number of fields that allows a complete determination of the spatial phases of all the fields from the demodulated beat signals.
Utilizing more fields may allow for more precise phase front determination at the expense of increased system complexity.

Finally, we briefly address the issue of linear independence of the signal wave vectors.
If all the $\phi_{ij}$'s vary uniformly over space (for instance, if they are launched from a single source antenna), then the $\theta^{\textrm{sp}}_{i}$'s register no change, and there is no way to determine the spatial variation in the phase pattern.

\subsection{Sensor array implementation}
\label{sec:sens-array-impl}
While this work used a single sensor taking data at multiple positions, this measurement can also be performed simultaneously using an array of sensors.
With such an array, each operating according to the principles described above, to first order we may consider them to be independent and non-interacting.
In general we expect the detection circuit of each sensor to have a slightly different phase response curve $\Theta^{\textrm{post}}(\omega)$.
In order to produce mutually consistent spatial phase readings between the multiple sensors, $\Theta^{\textrm{post}}$ must be independently calibrated for each sensor.
This can be measured easily using an amplitude-modulated field from each antenna individually and measuring the response at each sensor.